\input{epsf}

\documentstyle[prd,eqsecnum,aps,twocolumn]{revtex}

\catcode`\@=11

\def\maketitle2{\par 
\begingroup
\let\cite\@bylinecite
\def\thefootnote{\fnsymbol{footnote}}%
\twocolumn[\@maketitle2\vskip2pc]%
\thispagestyle{plain}\@thanks
\endgroup
\def\thefootnote{\arabic{footnote}}%
\setcounter{footnote}{0}%
\let\maketitle2\relax \let\@maketitle2\relax
\let\@thanks\relax \let\@authoraddress\relax \let\@title\relax
\let\@date\relax \let\thanks\relax \let\@abstract\relax 
\let\@pacs\relax}

\def\abstract#1{\gdef\@abstract{{\par 
\bgroup
\ifdim\prevdepth=-1000pt \prevdepth0pt\fi
\hsize\columnwidth
\dimen0=-\prevdepth \advance\dimen0 by17.5pt \nointerlineskip
\small\vrule width 0pt height\dimen0 \relax}{~~}#1\egroup}}

\def\pacs#1{\gdef\@pacs{{\par 
\bgroup
\hsize\columnwidth \parindent0pt
\ifdim\prevdepth=-1000pt \prevdepth0pt\fi
\dimen0=-\prevdepth \advance\dimen0 by20pt\nointerlineskip
\egroup} PACS numbers:~#1}}

\def\@maketitle2{
\@preprint
\@title
\ifdim\prevdepth=-1000pt \prevdepth0pt\fi
\@authoraddress
\@date
\begin{list}{}{\leftmargin=0.10753\textwidth \rightmargin=\leftmargin
\itemsep=1pc\partopsep=-1pc}
\item\@abstract
\item\@pacs
\end{list}
}

\catcode`\@=12

\begin{document} 

\newcommand {\vp} {\varphi}
\newcommand {\be} {\begin{equation}}
\newcommand {\ee} {\end{equation}}
\newcommand {\bea} {\begin{eqnarray}}
\newcommand {\eea} {\end{eqnarray}}
\preprint{LA-UR-97-1239}
\title{Compacton Solutions in a Class of Generalized Fifth Order Korteweg-de
Vries Equations } 
\author{ Fred Cooper$^1$, James M. Hyman$^2$ and Avinash
Khare$^3$} 
\address{$^1$Theoretical Division, MS B285, Los Alamos National
Laboratory, Los Alamos, NM 87545}   
\address{$^2$Theoretical Division, MS B284,
Los Alamos National Laboratory, Los Alamos, NM 87545}  
\address{$^3$Institute of
Physics, Sachivalaya Marg, Bhubaneswar 751005,India}  
\date{\today} 
\abstract
{We study  generalized Korteweg-de Vries  (KdV) equations derivable from the
Lagrangian: 
\[
 L(p,m,n,l)= \int~dx~ [ \frac{1}{2} \vp_{x} \vp_{t} + \alpha{
{(\vp_{x})^{p+2}} \over {(p+1)(p+2)}}- \beta (\vp_{x})^{m} (\vp_{xx})^{2}+
{\gamma \over 2} \vp_x^n \vp_{xx}^l \vp_{xxx}^2  ]. \] The usual field $u(x,t)$
of the generalized KdV equation is defined by $u(x,t) = \vp_{x}(x,t)$. The
equation of motion derived from this Lagrangian has solitary wave solutions of
both the usual (non-compact) and compact variety (``compactons"). For the
particular case that $p=m=n+l$, the solitary wave solutions have compact support
and the feature that their width  is independent of the amplitude.  We  discuss
the Hamiltonian structure of these theories and find  that mass, momentum, and
energy are conserved. We find in general that these are not completely
integrable systems. Numerical simulations show that an arbitrary compact
initial  wave packet whose width is wider than that of a compacton breaks up
into several compactons all having the same width. Upon scattering the
compactons found here behave similarly to those found in the equations of
Rosenau and Hyman, the scattering is almost elastic, with the left over wake
eventually turning into compacton-anticompacton pairs. Often there are two
different compacton solutions for a single set of parameters having the same
generic form $ A  \cos^r (d \xi)$, where $\xi= x-ct$ and $ -\pi/2  \leq d \xi
\leq \pi/2$.  When this is the case, the wider solution is stable, and this
solution is a minimum of the Hamiltonian. } \pacs{  03.40.Kf, 47.20.Ky, Nb,
52.35.Sb} \maketitle2 \narrowtext

\section{Introduction}

The observed stationary and dynamical patterns in nature are usually finite in
extent. However most equations that admit solitary wave solutions yield
solutions that are infinite in extent, although of a localized nature.
Therefore, the recently discovered solitary waves with compact support
(compactons) found by Rosenau and Hyman \cite{RH} represents a welcome
development. Rosenau and Hyman found in numerical experiments that the compact
solitary waves   on collision reemerge as compactons, with a tiny amount of
energy going into a zero mass ripple which eventually reemerges into a
compacton- anticompacton pair of small ($< 5\% $) amplitude.  The original
equation studied by  Rosenau and Hyman was \cite{RH}   \begin{equation}
 K(n,m) :   u_t + (u^m)_x + (u^n)_{xxx} = 0, \end{equation} For  $1 \le m=n
\le{3}$ this equation had solitary wave solutions of  the form $[\cos(a
\xi)]^{2/(m-1)}$, where $\xi= x-ct$ and $ -\pi/2  \leq a \xi \leq \pi/2$ . For
$m$=2,3 they obtained:

\begin{eqnarray}
    K(2,2): &  &  u_c = { {4c} \over {3}} \cos^2(\xi/4) \nonumber\\ 
    K(3,3): &  &  u_c = \left({3c} \over {2} \right) ^{1/2} \cos(\xi/3) .
\end{eqnarray}
 
Unlike the ordinary KdV equation the above equation is not derivable from a first
order Lagrangian except for $n=1$, and  did not possess the usual conservation
laws of energy, momentum,
 and mass that the KdV equation possessed.    Because of this, we   considered
earlier a different  generalization of the KdV equation based on a first order
Lagrangian formulation.  That is, we  considered \cite{bib:kdv1} \cite{bib:khare}

\begin{equation} L(l,p) = \int \left(  \frac{1}{2} \vp_{x} \vp_{t} + {
{(\vp_{x})^{l}} \over {l(l-1)}} - \alpha (\vp_{x})^{p} (\vp_{xx})^{2}  \right)
dx. \label{L} \end{equation}

This Lagrangian leads to a generalized  sequence of KdV equations of the form:
\begin{eqnarray} K^*(l,p):  u_t &+ & u^{l-2} u_x + \alpha (2u^p  u_{xxx} + 4p
u^{p-1} u_x u_{xx} \nonumber \\
 &+& p(p-1) u^{p-2} (u_x)^3 ) =0, \label{kstar} \end{eqnarray} where \be u(x,t)
= \vp_x(x,t) .\nonumber \ee These equations have the same terms as the equations
considered by Rosenau and Hyman, but the  relative weights of the terms are
different. In that study we showed that for $0 \leq p \leq 2$ and $l=p+2$, 
these models admit compacton solutions for which the width is independent of the
amplitude. Since higher order KdV equations also admit soliton solutions, it is
interesting to enquire whether higher order generalized KdV equations also admit
compacton solutions, and whether there are cases where the  width is independent
of the amplitude.  This is the first motivation for the present study. The second
motivation is to study the scattering properties of the compactons numerically
to see if they behave similarly to those found by Rosenau and Hyman. The
generalization which we will study in this paper is described by the Lagrangian:
\begin{eqnarray}
 L(p,m,n,l) &=& \int~dx~ [ \frac{1}{2} \vp_{x} \vp_{t} + \alpha{
{(\vp_{x})^{p+2}} \over {(p+1)(p+2)}} \nonumber \\ &-& \beta (\vp_{x})^{m}
(\vp_{xx})^{2}+ {\gamma \over 2} \vp_x^n \vp_{xx}^l \vp_{xxx}^2 
],\label{eq:genkdv2}  \end{eqnarray} which includes one extra term with higher
derivatives.  This Lagrangian Eq.~(\ref{eq:genkdv2}) generalizes both the usual
KdV Lagrangian and our previous generalized KdV Lagrangian, and shares with both
equations that because of the invariance of the action under time, and space
translations as well as the shift of the field by a constant $(\vp \rightarrow
\vp+a)$  that energy, momentum and mass are conserved by Noether's theorem. 

The rest of the paper is organized follows:  In section 2 we will discuss the
Hamiltonian structure of these classes of theories and determine the equation
for the  travelling wave solutions to (\ref{eq:genkdv2}). In section 3 we give
arguments based on a variational approach (see \cite{bib:kdv1} \cite{bib:khare})
as well as dimensional reasons for there being solutions with the width of the
solitary wave being independent of the amplitude and we determine the
connection between the energy and momentum of these solitary waves: namely $E
\propto Pc$. In particular we find that the condition for  having solitary waves
with width independent of amplitude is $p=m=n+l$. In section 4 we obtain exact 
compacton solutions of the form $A \cos^r d (x-ct)$ and verify that  the
relations among the global variables obtained using the variational approach are
exact.   In section 5 we give numerical results for the scattering of two
compactons, as well as for the breakup of an arbitrary wave packet into several
compactons.  In an appendix we show that most of the exact solutions as well as
the criteria for their stability could be obtained by assuming the exact compact
ansatz and minimizing the action on that class of functions. The unstable
solutions are  maxima of the effective Hamiltonian.

\section{Generalized KdV equation and Properties}

From the Lagrangian Eq.(~\ref{eq:genkdv2}) we obtain the generalized KdV
equation: \begin{eqnarray} u_t &+& {\alpha \over p+1} \partial_x u^{p+1} -\beta
m \partial_x(u^{m-1} u_x^2 ) + 2 \beta \partial_{xx} (u^m u_x ) \nonumber \\ &+&
{\gamma n \over 2} \partial_x(u^{n-1}u_x^lu_{xx}^2)-{\gamma l \over 2}
\partial_{xx}( u^n u_x^{l-1}u_{xx}^2) \nonumber \\
 &+ & \gamma \partial_{xxx}(u^nu_x^l u_{xx})=0, \label{eq:kdvnew} \end{eqnarray}
 as well as the conserved Hamiltonian: \begin{equation} H=\int dx [-\alpha{
{u^{p+2}} \over {(p+1)(p+2)}}  +\beta u^{m} u_{x}^{2}-{\gamma \over 2} u^n
u_{x}^l  u_{xx}^2  ], \label{eq:Hamiltonian} \end{equation} where $u(x,t)=
\vp_x(x,t)$. We notice that the Lagrangian given by Eq. (~\ref{eq:genkdv2}) is
invariant under the transformations (i) $\vp(x,t) \rightarrow \vp(x,t) + c_1$;
(ii)$ x \rightarrow x+ c_2$ and (iii) $t \rightarrow t+c_3,$ where $c_1$, $c_2$
and $c_3$ are constants. By a direct application of Noether's theorem this leads
to the three conservation laws of mass $M$, momentum $P$ and energy $E$, where
$E$ is given by Eq. (\ref{eq:Hamiltonian}) and $M$ and $P$ are given by
\begin{equation} M=\int u(x,t) dx; \hspace{.5in} P= {1 \over 2} \int u^2(x,t)
dx. \label{eq:P} \end{equation} The equation of motion Eq. (\ref{eq:kdvnew}) is
also invariant under the transformations as  given by \[ u \rightarrow ku;~~x
\rightarrow k^a x;~~t \rightarrow k^b t \] provided that \[ {n+l-p \over l+4} =
a= {m-p \over 2}; ~~ b=a-p.\] However, as seen below for the interesting
compacton case where $m=p=n+l$, these  equations are invariant under the more
restricted transformations: \begin{equation}
 u \rightarrow ku;~~x \rightarrow  x;~~t \rightarrow k^{-p}t . \label{eq:inv}
\end{equation} We shall see later that this condition is necessary so as to have
compactons whose width is independent of the amplitude. As an aside we would
like to point out that all the compactons so far discovered with width
independent of amplitude have the property that the field equations are
invariant under Eq.(~{\ref{eq:inv}). 

The canonical structure of these theories is similar to that found in
\cite{bib:kdv1} in that by postulating that the $u(x)$ satisfy the Poisson
bracket structure \cite{bib:Das} \begin{equation} \{u(x),u(y)\} = \partial_x
\delta (x-y) .\label{poiss1} \end{equation} we obtain that \begin{equation} u_t
= \partial_x  {{\delta H } \over {\delta u}} = \{u,H\} \end{equation} with $H$
being given by Eq.(~{\ref{eq:Hamiltonian}}). We also find that with our
definition of $P$ given by Eq.(~{\ref{eq:P}}), that $P$ is indeed the generator
of the space translations: \begin{equation} \{u(x,t), P \} = {\partial u \over
\partial x} . \label{genP} \end{equation} Since our equation is a generalization
of the equation discussed in  \cite{bib:kdv1}, one is also not able to show the
existence of a bi-Hamiltonian structure using the conserved momentum as a
possible second Hamiltonian as was done for the ordinary KdV equation. So on
these grounds one expects that our general Lagrangian does not correspond to an
exactly integrable system, except for the original KdV equation case. This will
be verified by our numerical results on scattering where there is some energy
going into compacton pair production following scattering.

\subsection{Equation for solitary waves } If we assume a solution to
(\ref{eq:kdvnew}) in the form of a travelling wave:
 \begin{equation} u(x,t)  =f(x-ct) \equiv f(y) , \end{equation} one obtains for
$f(y)$: \begin{eqnarray} c f_y &=& {\alpha \over p+1} \partial_y f^{p+1} -\beta
m \partial_y( f^{m-1}f_y^2)+ 2 \beta \partial_{yy} ( f^m f_y) \nonumber \\ &+&
{\gamma n \over 2} \partial_y(f^{n-1}f_y^lf_{yy}^2)-{\gamma l \over 2}
\partial_{yy}( f^n f_y^{l-1}f_{yy}^2) \nonumber \\
 &+& \gamma \partial_{yyy}(f^n f_y^lf_{yy}), \label{eq:kdvnew2} \end{eqnarray}
Integrating once we obtain: \begin{eqnarray} c f  &=& {\alpha \over p+1} 
f^{p+1}-  \beta m (f^{m-1}f_y^2) + 2 \beta  \partial_{y} (f^m f_y) \nonumber \\
&+& {\gamma n \over 2} ( f^{n-1}f_y^lf_{yy}^2)-{\gamma l \over 2}
\partial_{y}(f^n f_y^{l-1}f_{yy}^2) \nonumber \\ & +&\gamma \partial_{yy}(f^n
f_y^lf_{yy})+c_1, \label{eq:kdvnewint} \end{eqnarray} where $c_1$ is a constant
of integration. This equation needs several more integrations before a solution
in terms of quadrature is obtained, unlike the previous equation we studied
where two integrations were sufficient. Thus an explicit solution in terms of
quadratures is not available and one must use ``educated'' guess ansatze to find
exact solutions.  It is therefore quite useful to  look at a simple variational
approach to the solitary wave problem to understand some of the global features
of the solitary waves. Previous experience \cite{bib:kdv1}  \cite{compacton} 
with variational methods have shown that many of the global relations among
conserved quantities as well as conditions for the width to be independent of
the amplitude can be obtained from a simple variational ansatz.  Also whether
the  solitary wave is expected to be stable or not can be inferred from the fact
that the variational solution is a minimum of the action or just a saddle point
of the action.  Before addressing these questions, we just remind the reader
that at $\gamma=0$ the Lagrangian we are studying reduces to the previous
generalized KdV problem we studied \cite{bib:kdv1} so that these equations
include all the KdV solitons and compactons that we discussed earlier as a
special case. We will find exact solutions of Eq. (\ref{eq:kdvnewint}) by
inserting the ansatz: \begin{equation} f(y) = A \cos^r (dy) \end{equation} and
determining the parameters $A$, $r$, and $d$ by a consistency argument.

\section{Variational Approach}

Our time-dependent variational approach for studying solitary waves is based on
the principle of least Action.   In  previous work \cite{bib:kdv1} \cite{CSLS1}
\cite{CSLS2}, we introduced a post-Gaussian variational  approximation, a
continuous family of trial variational functions more general than  Gaussians,
which can still be treated analytically. We assumed a variational ansatz of the 
form  \[ u_v(x,t) =  A(t) \exp \left[-b (t) \left| x-q(t) \right|^{2 s}  \right]
\] The variational parameters have a simple interpretation in terms of
expectation  values with respect to the ``probability'' $\bar{P}$ \be \bar{P}
(x,t) = \frac{\left[u_{v}(x,t)\right]^{2}}{2 P}, \ee where the conserved
momentum $P$ is defined as above

\be P \equiv {1 \over 2}\int \left[u_{v}(x,t)\right]^{2} dx \label{M}. \ee We
have that
  \[ q(t) = \langle x \rangle\]. \be G_{2s} \equiv \langle |x-q(t)|^{2s} \rangle
= \frac{1}{4s b}. \ee and \be A(t) =  \frac{P ^{1/2}
(2b)^{1/4s}}{\left[\Gamma\left(\frac{1}{2s} +  1\right)\right]^{1/2}}. \ee 
 Extremizing the effective action for the trial wave function $u_v$ leads to 
Lagrange's equations  for the variational parameters.  We  find that for all
values of the parameters ($l, m, n, p$) the dynamics of the variational
parameters lead to solitary waves moving with constant velocity and constant
width. However the solutions found are often maxima or saddle points of  the
action.  We will find that when this happens the exact
solitary wave solution is unstable. 
 For the special case of $p=m=n+l$ we will show below that the width of the
soliton is independent of the amplitude and velocity.  For that case we also
obtain the relationship: \[ E= -{2 c \over p+2} P\] We will find that the exact
solitary wave solutions satisfy this relationship as long as the integration
constant $c_1$  of Eq.~(\ref{eq:kdvnewint})is zero.

The starting point for the variational calculation is the action

\be \Gamma = \int L dt, \label{action} \ee where $L$ is given by (\ref{L}).

Inserting the trial wave function $u_v$ we obtain:

\begin{equation} \Gamma(q,\beta,P,s) =  \int dt  [ -P  \dot{q}-  H_{eff} ];
\end{equation} where $H_{eff}$ is the Hamiltonian evaluated using the variational
wave function $u_v$.  We find \bea H_{eff}  & = &  C_{1}(s)  b^{p/4s} P^{(p+2)/2}
+ C_{2}(s) b^{(m+4)/4s} P^{(m+2)/2}   \nonumber \\ & + & C_3(s) b^{(n+3l+8)/4s}
P^{(n+l+2)/2}. \eea where $C_1$, $C_2$ and $C_3$ are functions of
$s,\alpha,\beta,\gamma$ but independent of $P,b$.  We notice that there is no
momentum conjugate to $b$.
 We eliminate the variable of constraint $b$ (using $\delta\Gamma/ \delta b =
0$) and  obtain the equation:

\bea 0  & = & {p \over 4s} C_{1}(s)  d^p P^{(p+2)/2} + {m+4 \over 4s} C_{2}(s)
d^{(m+4) P^{(m+2)/2} }  \nonumber \\ & + & {(n+3l+8) \over 4s}C_3(s)
d^{(n+3l+8)} P^{(n+l+2)/2},\label{beta} \eea where  \be d = b^{1 \over 4s}.
\label{d} \ee

From Eq. (\ref{beta})  we see that when \be p=m=n+l \ee the momentum $P$ factors
out of the equation, so that the width of the soliton which depends on  $b$ 
does not depend on $P$ and thus is independent of the amplitude or velocity. 
This special case is precisely the case when the exact  solution is a compacton
whose width is independent of amplitude as we shall show below.  From Lagrange's
equation $ \delta \Gamma / \delta P =0 $ we have that \begin{equation} -{\delta 
H_{eff} \over \delta P} = \dot{q} = c.  \end{equation} When $p=m=n+l$ this
yields the relationship: \begin{equation} E=- {2 c \over p+2} P, \label{eq:rel}
\end{equation} which we will verify is true for the exact compacton solutions
when $c_1=0$. For this special case of $p=m=n+l$ the functions $C_i$ are given
by:
 \[ C_1= -N \alpha{\Gamma({1 \over 2s}) \over (p+1)(p+2) s (p+2)^{1 \over
2s}} \]
 \[ C_2 = 4 N \beta  s ~\Gamma( 2-{1\over 2s}) ~(p+2)^{{1 \over 2s}-2 }
 \]
\begin{eqnarray}
&& C_3 =-(2s)^{l+1}(-1)^l N \gamma  (p+2)^ {{l+3 \over
2s}-4-l}  ~\Gamma(l+2-{l+3 \over 2s}) \nonumber \\
 && (2s-1) [p^2(2s-1)+2p(1+l-2ls)+l^2(2s-1)\nonumber \\
 &+& 2l(s-1)+4s-1] 
\end{eqnarray}
 where $N= 2^{p+2 \over 4s}/ [\Gamma(1+ {1 \over 2s})]^{p+2
\over2}.$ 
 
For the special case that $p=m=n+l$ our procedure for determining the optimal
trial wave function is very straight forward. The effective  Hamiltonian
simplifies to (from now on we drop the subscript on $H_{eff}$): 
\begin{equation}
H(d,s)= P^{{p+2 \over 2}} d^p [C_1(s) +d^4 C_2(s)+d^{2l+8} C_3(s)]
\end{equation}
Minimizing with respect to the width parameter $d$, we obtain: 
\begin{equation}
pC_1 +(p+4) d^4 C_2 + (p+2l+8) d^{2l+8} C_3=0 
\end{equation} 
For $l=0$, one can
analytically solve this for $d(s)$ and one usually finds that there are 2 real
solution in the vicinity of  $ s = 1$  which is the Gaussian trial wave
function.  Inserting these 2 solutions ($d_i(s)$) into $H$ we obtain an
analytic expression of the form  \[ H_i(s) = H(d_i(s),s) \] Imposing $ dH_i/ds
=0$, we find that the two solutions $ \{d_i(s_i),s_i \}$ (i=1,2) have the
property that one is a saddle of $H$ in the $\{d,s\}$ plane and one is a
minimum. We find in our numerical simulation that only the solutions that is a
 minimum is stable.  In general it is easy to find the extrema of $H$ in
the $\{d,s\}$ plane graphically by making a 3 dimensional plot of $H$.
  What is convenient about the class of trial wave functions discussed here, is
that one can explicitly determine the effective action for all values of
$p,l,m,n$ and get the general features of the results for all values of the
parameters.  In the appendix we will use the  form of the  exact  compacton wave
function in the variational approach.  By doing so we will recover most of the
exact compacton solutions that we find in the next section. We also obtain the 
result that when there are multiple solutions the stable one is a minimum of the
effective Hamiltonian for the exact shape function.

\section{Exact Compacton Solutions}

Motivated by our
variational results we will concentrate on the particular case where $p=m=n+l \neq 0$. 
which corresponds to Compactons whose width is independent of amplitude. 
Assuming a solution of the form 
\begin{eqnarray}
 u(x,t)&=& A \cos^r[d(x-ct)];
{-\pi \over2} \leq d(x-ct) \leq {\pi \over 2}\nonumber \\ u(x,t)&=&0; ~~~ |d
(x-ct) | > {\pi \over 2}, 
\end{eqnarray}
we look for consistent solutions for
$A$, $r$,$c$ and $d$ in terms of $\alpha$, $\beta$ and $\gamma$ and $c_1$ which
are real. We check whether there is an analogous variational
solution of the post-Gaussian variety corresponding to a minimum and not just a
saddle point or maximum of the action. We  also show in the appendix that if we a
assume a trial wave function of the exact type: \[u(x,t)= A \cos^r[d(t) (x-
q(t)]\] the exact solutions are either maxima or minima of the effective
Hamiltonian as a function of $d$. We find that when there is a unique solution,
then it is a minimum of the Energy. When there are two solution, the
wider one is stable and is also a minimum of the Energy. The narrower one
is unstable and is a  maximum of the Energy.

Now let us look at some special cases.  
\subsection{$p=m=n$ and $l=0$ case}

For the case $p=m=n$ and $l=0$ it is possible to find a general class of 
solutions for arbitrary $p$. Inserting a trial solution of the form:
\begin{equation} u(y)= A ~ \cos^r(dy) \end{equation} into
Equation(\ref{eq:kdvnewint}) we obtain the consistency  equation:
\begin{eqnarray}
 0&&= -3c_1 + 2Ac{x^r} \nonumber \\ && + {A^{1 + p}}{x^{-4 + r + rp}}\left( 1 -
r \right) r{d^4}\gamma  \times \nonumber \\ &&\left( 12 - 10r + 2 {r^2} - 11rp +
5{r^2}p + 
    2{r^2}{p^2} \right)  \nonumber \\ && + 2{A^{1 + p}} {x^{-2 + r + rp}}r{d^2}
  \left( -2 + 2r + rp \right) \times  \nonumber \\
 && \left( -\beta  + 2{d^2}\gamma  - 
    2r{d^2}\gamma  + {r^2}{d^2}\gamma  - 
    r{d^2}\gamma p + 2{r^2}{d^2}\gamma p
     \right) \nonumber \\ &&+ {A^{1 + p}}{r^2}{d^2}\left( 2 + p \right) 
  \left( 2\beta  - {r^2}{d^2}\gamma  - 
    2{r^2}{d^2}\gamma p \right) {x^{r + rp}} \nonumber \\ &&- {2\over {1 + p}}
\,{A^{1 + p}}\,\alpha \,x^{r + r\,p}\label{eq:big} \end{eqnarray}

Here $x= \cos(dy)$.  All the powers of $x$ must have zero coefficient for the
trial solution to be an actual solution. This leads to various conditions
depending on the values of $r$ and $p$. If $ r(1+p)=4$ then there can also be
solutions with $c_1$ being nonzero. First let us consider the case when $c_1=0$.
In that case for consistency we need either $rp=2$ or $rp=4$.
\subsubsection{rp=2} When $ rp=2$ Eq. \ref{eq:big} reduces to the conditions:
\begin{eqnarray} 0 &&= \gamma \,
  {{\left( -1 + r \right) }^2}\,r\,\left( 1 + r \right), \nonumber \\
 A^{2/r}&& =  -\alpha  + 4\,\beta \,{d^2} + 6\,r\,\beta \,{d^2} + 
  2\,{r^2}\,\beta \,{d^2} - 8\,r\,{d^4}\,\gamma  \nonumber \\ && -
14\,{r^2}\,{d^4}\,\gamma  - 7\,{r^3}\,{d^4}\,\gamma  - 
  {r^4}\,{d^4}\,\gamma , \nonumber \\ 0 &&= -\alpha  + 4\,\beta \,{d^2} +
6\,\beta \,{d^2}\,r - 
  8\,{d^4}\,\gamma \,r + 2\,\beta \,{d^2}\,{r^2}\nonumber \\ && -
14\,{d^4}\,\gamma \,{r^2} - 7\,{d^4}\,\gamma \,{r^3} - 
  {d^4}\,\gamma \,{r^4}. \end{eqnarray} The first condition tells us that either
$\gamma=0$ or $r=1$.

When $\gamma=0$, we get the solution we found in our earlier work
\cite{bib:khare}: Namely: \[ d^2 = {{\alpha \,{p^2}}\over 
   {4\,\beta \,\left( 1 + p \right) \,
     \left( 2 + p \right) }} \] \[ A^p= \{ {{c\,\left( 1 + p \right) \,\left( 2
+ p \right) }\over
      {2\,\alpha }}\} \] When $\gamma \neq 0$ we instead get the solution 
$r=1$  ($p=2$)  and \[ A^2= {c\over {2\,\beta \,{d^2} - 6\,{d^4}\,\gamma }}\] and
two possible solutions for the width: \[ d^2= {{12\,\beta  \pm 
{\sqrt{144\,{{\beta }^2} - 
         120\,\alpha \,\gamma }}}\over {60\,\gamma }} \] Which means we also can
write the equation for $A^2$ as \[ c= {A^2 \over 5} (\alpha- 2 \beta d^2) \]
Thus when $ \gamma \neq 0$ one only gets a solution when $p=2$. A particular
case of this solution  is:
 $\alpha=6$, $\gamma=3$ and $\beta =4$. Then there are two solutions   having 
$d^2=1/3$ or $d^2=1/5$. The first is \begin{equation} u=\sqrt{3 c/2} \cos ({x-
ct \over \sqrt{3}}) \end{equation}

The conserved quantities for this solution are \[~M= 3 \sqrt{2 c}; P= {3 \over
8} \sqrt{3} c \pi; E=  -{3 \over 16} \sqrt{3} c^2 \pi \] Thus we find that the
relationship: \begin{equation} {E \over P} = -{2 c \over p+2} = -{c \over 2}
\end{equation} that we derived from our variational approach is exact here.

The second solution : \begin{equation} u= \sqrt{ 25 c \over 22} \cos {y \over
\sqrt{5}} \end{equation} has as its conserved quantities:

\[~M= 5 \sqrt{{10 c \over 11}}; P= {25 \sqrt{5}  c \pi \over 88} ;
 E=  -{25 \sqrt{5} c^2 \pi \over 176} \] Thus again we find that the
relationship: \begin{equation} {E \over P} = - {2 c \over p+2} = - {c \over 2}
\end{equation} is exact.  

To understand whether these solutions are stable solutions we use
the variational principle using   a trial wave function
of compact  form: \begin{equation} u= A \cos [dy] ;~~  - {\pi \over 2} \leq dy
\leq {\pi \over 2}; ~~ y=x- q(t) \end{equation} where the relationship \[ A^2 =
{4 d P \over \pi}
 \] pertains.  (See the appendix for details)  If we insert this into the
action, we obtain  for the reduced Hamiltonian  (where $L = -P \dot{q} - H[P]
$),  using the above values of $\alpha$,$\beta$ and $\gamma$: \begin{equation}
H=  { P^2 \over \pi}  (-9 d^5 +8  d ^3 - 3  d) \end{equation} From  Lagrange's
equations  $ {\partial H \over \partial P} = - \dot{q}$ we obtain a $ 2 E/P =  
-\dot{q} = -c $.  Minimizing the energy with respect to $d$ gives back the exact
result that $d^2=1/5$ which is a  minimum of the energy and $d^2=1/3$
which is a  maximum of the energy.  This suggests that the narrower
compacton with $d^2 = 1/3$ is unstable. We
 have confirmed  this numerically.  We have also shown numerically (see below)
that if we start with  initial compact data which is wider than the compacton
with $d^2=1/5$ it breaks up into a number of these compactons.
\subsubsection{rp=4 case}

Let us now consider the case $ r={4 \over p}$, with $c_1=0$.  When $\gamma=0$
one of the consistency conditions   $ A c=0$ can only be satisfied for static
solutions.  When $\gamma \neq 0$ the consistency conditions leads to :
\begin{eqnarray} d^2 &&=  {\beta \over \gamma (r^2+6 r-2)}, \nonumber \\
A^p&=&{c\over \left( r - 1 \right) \,r^2\, \left( 5 + r \right) \, d^4\,\gamma }
\end{eqnarray} with the parameters also obeying the constraint: \begin{equation}
\alpha={{\left( 2 + r \right) \,\left( 4 + r \right) \,
 \left( -4 + 4\,r + {r^2} \right) \,{{\beta }^2}}\over {{{\left( -2 + 6\,r +
{r^2} \right) }^2}\,\gamma }} \end{equation} Let us look at  2 examples from
this class of solutions: For the case \\ (i) $p=m=n=1; l=0$ \\
 the solution  is of the form: \[ u= A \cos ^4(dy)  \] Choosing for example
$\gamma=1/38 $ and $\beta=1$, the above relations yields $ d=1$, $A= 19 c/ 216
$  and $\alpha=672/19 $. Thus the solution is \begin{equation}
 u = {19 c \over 216} \cos^4 (x-ct) \label{eq:sol1} \end{equation} For this
solution we have that the global quantities are: \[  M={19 c \pi \over 576};~~P=
{12635 c^2 \pi \over 11943936};~~E= -{12635 c^3 \pi \over 17915904} \] So that $
E= - {2 \over 3} c P $.
  We show in the appendix that this  solution can also be found by minimizing
the effective action using the compacton ansatz. In that case one also gets
another solution with  $d^2= {77 \over 18}$, which is the  maximum of
the Hamiltonian, with $d^2=1$ being a  minimum as a function of $d$. 
The Hamiltonian as a function of the width $d$ for the above choice of parameters
is given by: \[ H= {{192\,{\sqrt{{ d}}}\,
     \left( -77 + 19\,{{{ d}}^2} - 
       2\,{{{ d}}^4} \right) \,
     {{{P}}^{{3\over 2}}}}\over 
   {95\,{\sqrt{35\,\pi }}}}.\] 

For the  case  \\ (ii) $p=m=n=2$,$l=0$ \\ In this case   we obtain the
conditions:  \begin{equation}
 r= 2; ~~ d^2 = {\beta \over 14 \gamma};~~ A^2 = {c \over 28 d^4 \gamma} 
\end{equation} Choosing $\beta=1$, $\gamma= {1 \over 14}$, $\alpha = {96 \over
7}$, we obtain $c= 2 A^2$ and the  solution is  \begin{equation} u= \sqrt{{c
\over 2}} \cos ^2 (x - c t). \end{equation} For this choice of parameters we
obtain for the conserved quantities: \[  M= {\pi \sqrt{c} \over 2 \sqrt{2}};~P =
{3 \pi c \over 32}; E=-{3 \pi c^2 \over 64}. \] We notice that we obtain the
relationship: $E/P = -2c /(p+2)$ which we obtained by the variational method.
For these parameters, if we use the compacton form as the trial wave function,
and use the relation: \[A^2= {16 d P \over 3 \pi}, \] then the effective 
Hamiltonian for the variational parameter $d$ is \begin{equation} H= -{8 P^2
\over 9 \pi} (10 d -5 d^3 +  d^5) \end{equation} This Hamiltonian has 2
stationary points as a function of $d$, $d=1$ ( minimum) and
$d=\sqrt{2}$ (maximum) The second solution is not a solution to the
equation of motion.

Next, let us  consider a particular special solution for the case  $ c_1 \neq
0$--  $p=m=n=1$,$l=0$. \\ For this case we  have $r=2$. Assuming a solution of
the form  \[ u(y) = A \cos^2(dy) \] we get the consistency equations:
\begin{equation} c_1= -6 A^2 d^4 \gamma . \end{equation} \begin{equation} d^2=
{1 \over 12 \gamma} [\beta \pm (\beta^2- \alpha \gamma)^{1/2}] \end{equation}
\begin{equation}
    A = {c \over 8 d^2 (\beta - 8  \gamma  d^2)}.   \end{equation}

For the special choice of $ \alpha =5$, $\beta = 3 $ and $\gamma = 1$, there are
two real solutions for $d^2$,corresponding to $d^2= 1/12$ and $d^2= 5/12$.  The 
first solution which is stable and which we will discuss further in our section
on numerical simulations is : \begin{equation} u = {9 c \over 14} \cos^2
({x-ct\over \sqrt{12}}). \label{eq:sol2} \end{equation} This solution has for
its constants of motion:
 \[ M= {9\sqrt{3}\pi c  \over 14 };~P = {243 \pi c^2 \sqrt{3} \over 1568};
E={{2349\,{c^3}\,\pi }\over {21952}}. \] Thus we obtain: $${E \over P} = {29 c
\over 42}$$ This shows  a failure of the relationship \ref{eq:rel}. 

The second solution  which we found to be numerically unstable is 
\begin{equation} u= - {9 c \over 10} \cos^2 [\sqrt{{5 \over 12}} (x-ct)]; ~ c >
0. \label{eq:sol3} \end{equation} The conserved quantities $E$, $M$, $P$ are
given by 

\begin{equation} M= - {9 \pi c \sqrt{3} \over 10 \sqrt{5}};~P = {243 \pi c^2
\sqrt{3} \over 800 \sqrt{5}};~ E=-{81 \pi c^3 \sqrt{3} \over 1600 \sqrt{5}}.
\end{equation} Now we find that $E/P = - c/6 \neq  - 2c /(p+2)$.

The  special solutions with $ c_1 \neq 0$ are not obtainable from a variational
calculation but these are a very restricted class of  solutions.

\subsection{Case $p=m=l,n=0$}

For this case inserting a trial solution of the form  $u= A \cos^r dy$ usually
leads to an overdetermined set of equations. 

For example for $p=1, r=2$ we obtain the conditions ($ x= \cos dy $):
\begin{eqnarray} 0=&& - c_1 - 48 {A^2} {d^5} \gamma  \sin (d y) x + 
  \left( A c - 8 {A^2} \beta  {d^2} \right)  
   {x^2}  \nonumber \\ && + 96 {A^2} {d^5} \gamma  \sin (d y) 
   {x^3} + \left(  -{1 \over 2} {A^2} \alpha   
        + 12 {A^2} \beta  {d^2} \right) {x^4} \end{eqnarray} which has only a
trivial solution.

For $p=2$ the situation is simpler and one obtains when $r=1$ the two relations:
\[ A^2= {c\over {2\,\beta \,{d^2} + 5\,{d^6}\,\gamma }} \] and  \[ 0= 18 d^6
\gamma+ 12 d^2 \beta - \alpha \]

These equations can have 2 or even 3 positive solutions for $d^2$. One
particular case is $\alpha=216$, $\beta=21$, $\gamma=-2$. In that case we get
two positive solutions for $d^2$ namely $d=1$ leading to : (i)\begin{equation}
 u= {\sqrt{c/2} \over 4} \cos(x-ct)  \end{equation} where the constants of
motion are: \[ M= {1 \over 2} \sqrt{c/2}; P={c \pi \over 128}; E= -{\pi c^2
\over 256} \] and $d^2 = 2$ giving the solution: (ii) \begin{equation} u=
{\sqrt{c} \over 2} \cos {\sqrt{2}(x-ct) }  \end{equation} and we obtain for the
conserved quantities: \[ M=  \sqrt{c/2}; P={c \pi \over 16 \sqrt{2} }; E= -{\pi
c^2 \over 32 \sqrt{2} } \] Thus both these solutions again obey the relation:
$E/P= - 2c/(p+2)$.  The second solution with $d^2=2$  turns out  to be
numerically unstable.

For $p=2$ and $r=2$ we get   relations: \begin{eqnarray}  c_1&& = 8 A^3 d^6
\gamma \nonumber \\
  c&&  - - 208 A^2 d^6 \gamma \nonumber \\ &&  -\alpha+48 \beta d^2 + 1152 d^6
\gamma =0  \end{eqnarray}

as well as one  constraints among the parameters  \[ \beta = - 48 d^4 \gamma \]
Eliminating the constraint, we obtain for the width: \begin{equation} d^6 = -
{\alpha \over 1152 \gamma} \end{equation} As an example if we choose $\gamma=-3$
and $\alpha = 3456$ then we have $d=1$ and for our solution: \[ u = {1 \over 4}
({c \over 39})^{1/2} \cos^2 (x-ct) \]

We have not exhausted all possible solutions for this case, but the method for
finding them should be clear to the reader by now.

\subsection{Some Other More General Cases} When $l+n=p=m$, we instead have one
parameter family of solutions depending on the velocity $c$. That is for fixed
$\alpha$ $\beta$ and $\gamma$ there is a solution of different amplitude for
different velocities $c$. In some special cases there is  the possibility for
two different solutions with the same value of $c$. However, in general  for a
given  $\alpha$ $\beta$ and $\gamma$  there is only one solution with a fixed
velocity $c$. To illustrate this fact, let us consider the case \[ p=n;~ m=l=0\]
Inserting a trial solution of the form \begin{equation} u= A \cos^{2/p}[ d
(x-ct)]; ~ p \neq 2 \end{equation} into Equation(\ref{eq:kdvnewint})  we obtain
for example if $\gamma=1$,$\beta=1$,  and $\alpha = -1/2$ the conditions: \[
d={p \over 2}[(p+1) (2 p^2+4p+3)]^{-1/4} \] \[  A^p ={ 8 \beta \over (4-p^2)}
[(p+1) (2 p^2+4p+3)]^{1/2} \] \begin{equation} c={ 2 \beta (p^2 + 8p +4) \over
(4-p^2) \sqrt{(p+1) (2p^2+4p+3)}}. \end{equation} Choosing for example $p=n=1$,
$l=m=0$, $\gamma=1$ and $\alpha=-1/2$ we get the single solution:
\begin{equation}  u = 8 \sqrt{2} \cos^2 [{ (x-{13 \sqrt{2} \over 9 }t) \over 2
(18)^{1/4}}] \label{eq:sol3a} \end{equation} The constants of motion for this
case are: \[ M= 8 \pi  (2)^{3/4}\sqrt{3}; P = 48 \pi (2)^{1/4}\sqrt{3}; E= -{400
\pi (2)^{3/4}\over 3 \sqrt{3}} \] So that $ E/P = -{25 c \over 39}$ For the
special case when $p=n=2$,one obtains \[u = A \cos dy \] When $m=l=0$  we find 
\[ d^4 = -{ \alpha \over 30 \gamma}; A^2 = {150 \gamma \over \alpha^2}(c+ {\alpha
\beta \over 15 \gamma}). \] Choosing further $\beta=\alpha=1$; $\gamma=-1/30$,
we obtain \[ u= \sqrt{10-c}~ \cos y \] So again for this special case we get a
continuous family of solutions as long as $ c < 10$. The constants of motion are
now: \[ M= 2 \sqrt{10-c}; P = {\pi \over 4} (10-c); E= {\pi \over 40} (10-c)
(30-c). \]

For any particular value of $p,~m,~n,~l$ one can always find the consistency
condition on the parameters $\alpha$, $\beta$, $\gamma$, $A$, $d$, $q$ and $c$
so that solitary wave solutions of the form $ A \cos^q[dy]$ exist.  However we
do not have a simple expression for these parameters for all $p,~m,~n,~l$, and
have instead in the above looked at some simple cases.

\section{Numerical study of the Generalized KdV equation}

\subsection{Numerical Method}

In our calculations, we approximated the spatial derivatives  with a
pseudo-spectral  method using the discrete  Fourier transform (DFT).   The
equations were  integrated  in time with a variable order, variable timestep
Adams-Bashford- Moulton method.   The numerical errors were monitored by 
varying the number of discrete Fourier modes between  128 and 512 and varying
the estimated time error per unit step between  $10^{-6}$ and  $10^{-9}$ to
insure that the solutions were well converged.   Mass and momentum were
conserved to an accuracy of at least $10^{-6}$  and the Hamiltonian was
conserved to an accuracy of better than  $10^{-2}$ in all of the calculations.  

The numerical approximation must respect the  delicate balance between the
nonlinear numerical dispersion terms  in the equation.  For example, when the
third term in (2.1)  is expanded, it has a diffusion-like term  $2 \beta m
(u^{m-1}u_{x} )u_{xx}$.   On the trailing edge of the  compacton $u_{x} > 0$ and
this term acts  like a destabilizing backward  diffusion operator.  The solution
would be  unstable if it were not for the  stabilizing nonlinear dispersion.  
This balance is easily lost in numerical  approximation if the aliasing, due  to
the nonlinearities, is not handled  carefully.  To identify numerical  artifacts
due to aliasing and other  discrete affects, we  solved the equations with the
nonlinear  terms expanded in different formulations.  For example, we compared
the solutions of  the equations when  they were differenced in  conservation
form and non  conservation form to identify possible  numerical inaccuracies. 

Also, the lack of smoothness at the edge of the compacton reduces the  spectral
method to first order near the edge and introduces dispersive errors  into the
calculation.  To reduce these errors and the errors due to aliasing, we 
filtered the time derivatives by explicitly adding an artificial dissipation
term to  the equations.  This term was defined in Fourier space to approximate
the  effects of linear second order dissipation ($\Delta x u_{xx}$) on the top
1/3 of  the Fourier modes, have no effect on the lower 1/3 of the modes and used
a  linear transition between the two regions.  

Most of the calculations  shown here solved the conservative  form of the
equations with 128 DFT modes and  a time error of $10^{-8}$ per unit time.
\subsection{Numerical Investigations}
For the original Rosenau and Hyman compactons, numerical investigations showed
some remarkable properties-- namely that whatever initial compact data was
given, it eventually evolved into compactons. When two compactons scattered any
energy not in the original pair of compactons emerged as compacton-anticompacton
pairs. We will find that the compactons of this fifth order
generalized KdV equation have similar properties to those previously found in
the studies of Rosenau and Hyman on their third order generalized KdV equation.
 
  The first generic feature of these equations is that arbitrary initial
compact data, as long as the width of the packet is larger than that of
the compacton evolves into several compactons with the number depending
on the initial energy. 
We show this for two different cases.  The first case is related
to the compacton of Eq. (4.20).  We start of with an initial 
pulse which is four times the width of the compacton and watch it evolve.
This is shown in Fig. 1.  
\epsfxsize=3.5in
\centerline{\epsffile{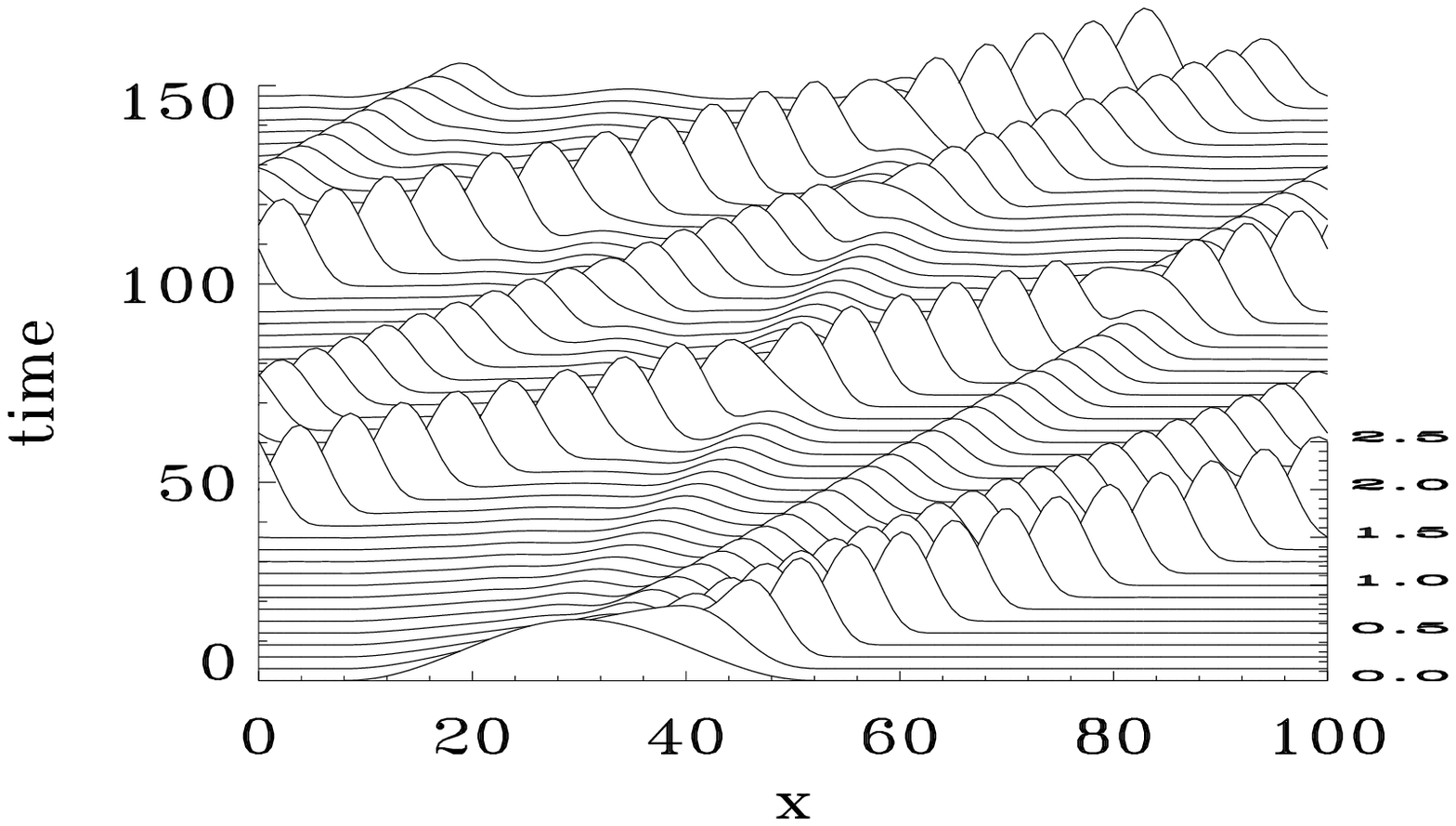}}
\protect\label{01b-w4-st}
\noindent
Fig. 1.   Pulse with an initial  width four times that of the
 compacton of Eq. (4.20)  pertaining to the parameters   
$p=m=n=1, l=0$ and 
$\alpha = 5, \beta = 3, \gamma = 1$ 
namely:  
$u_0 = {9 \over 14} \cos ^2 ( {x-30 \over 4 \sqrt{12} })$.
The initial wide pulse breaks 
into compactons that 
collide elastically.  Note the 
phase shift of the slower 
pulse after colliding with a faster, 
higher compacton.  \\

In figures 2 and 3 we show the same phenomena
for the compacton system described by Eq. (4.7), again starting from
initial data wide compared to the compacton solution. 

\epsfxsize=3.5in
\centerline{\epsffile{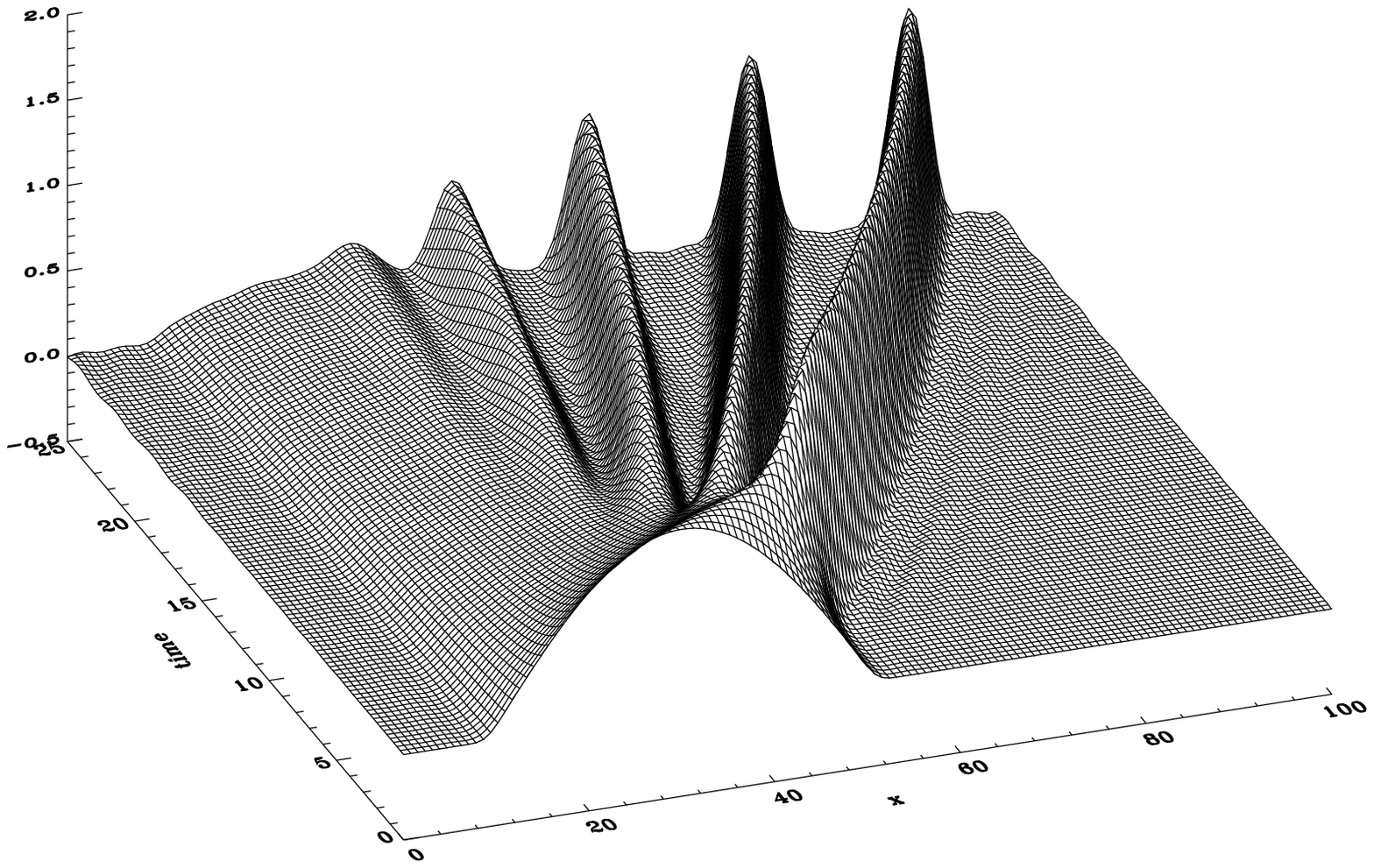}}
\protect\label{02b-w6-su}
\noindent
 Fig. 2.
Similar situation as in  Fig. 1 for the parameters
$p=m=n=2, l=0$ and 
$\alpha = 6, \beta = 4, \gamma = 3$ relevant to Eq. (4.7).
An initial compact wave (solid line)
$u_0 = \sqrt {\frac{25}{22} } \cos ( \frac{x-30}{6})$  wider than a compacton
with  breaks into a string of compactons with the shape 
$A \cos ({x-ct \over \sqrt{5}})$.

\epsfxsize=4.5in
\centerline{\epsffile{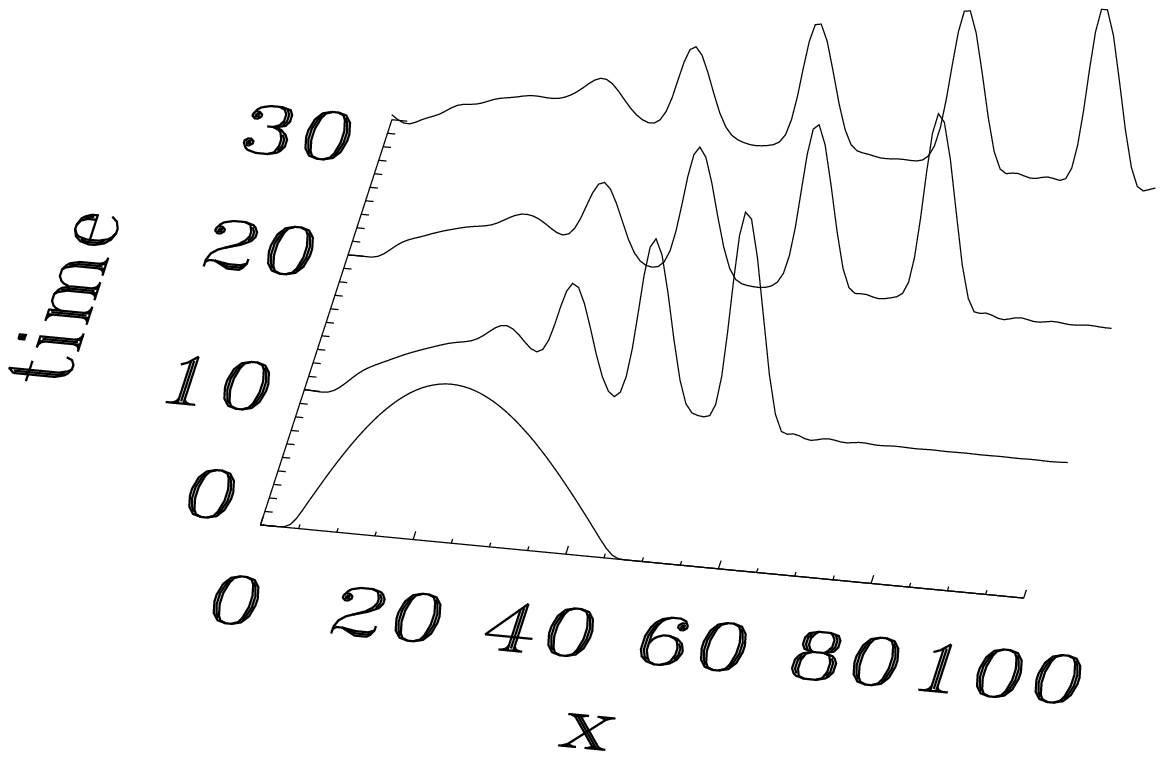}}
\protect\label{02b-w6-st}
\noindent
Fig. 3.
Same  compact wave break up as in Fig.2 displayed differently. \\

In figures 4, and 5 we show similar features of the breakup of 
a compact wave for the  compacton Eq. (4.13).
\epsfxsize=3.5in
\centerline{\epsffile{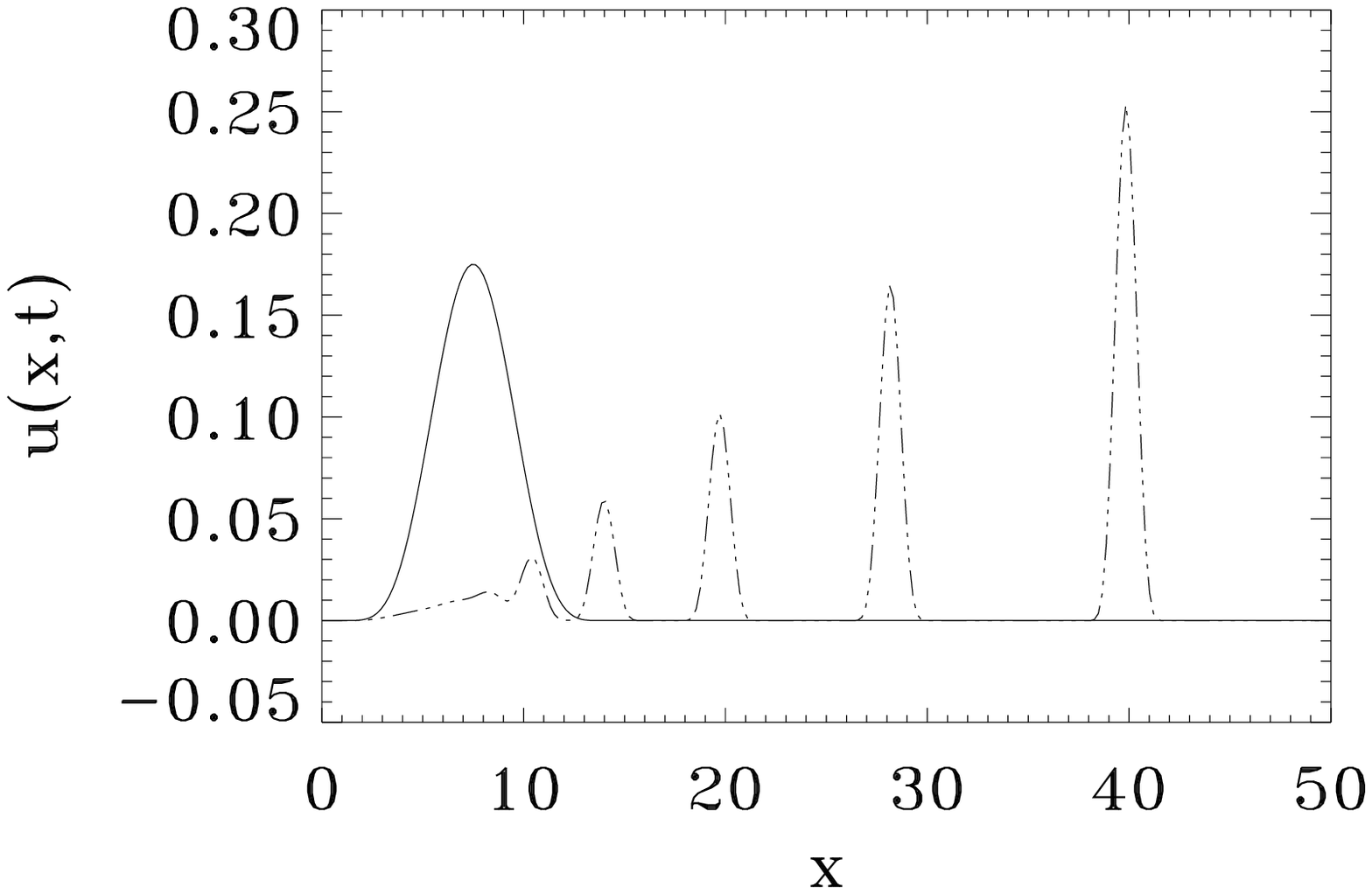}}
\protect\label{01c-w4-ol}
\noindent
Fig. 4 Break up of a compact wave with four times the width of the
compacton: $p=m=n=1, l=0$ and 
$\alpha = \frac{672}{19}, \beta = 1, 
\gamma = \frac{1}{38}$.
An initial compact wave (solid line)
$u_0 = 2 \frac{19}{216} \cos^4 (\frac{x-7.5}{4})$
breaks into a string of compactons with the shape 
$A \cos^4 (x-ct)$ by time $t=10$ (dashed line).

\epsfxsize=3.5in
\centerline{\epsffile{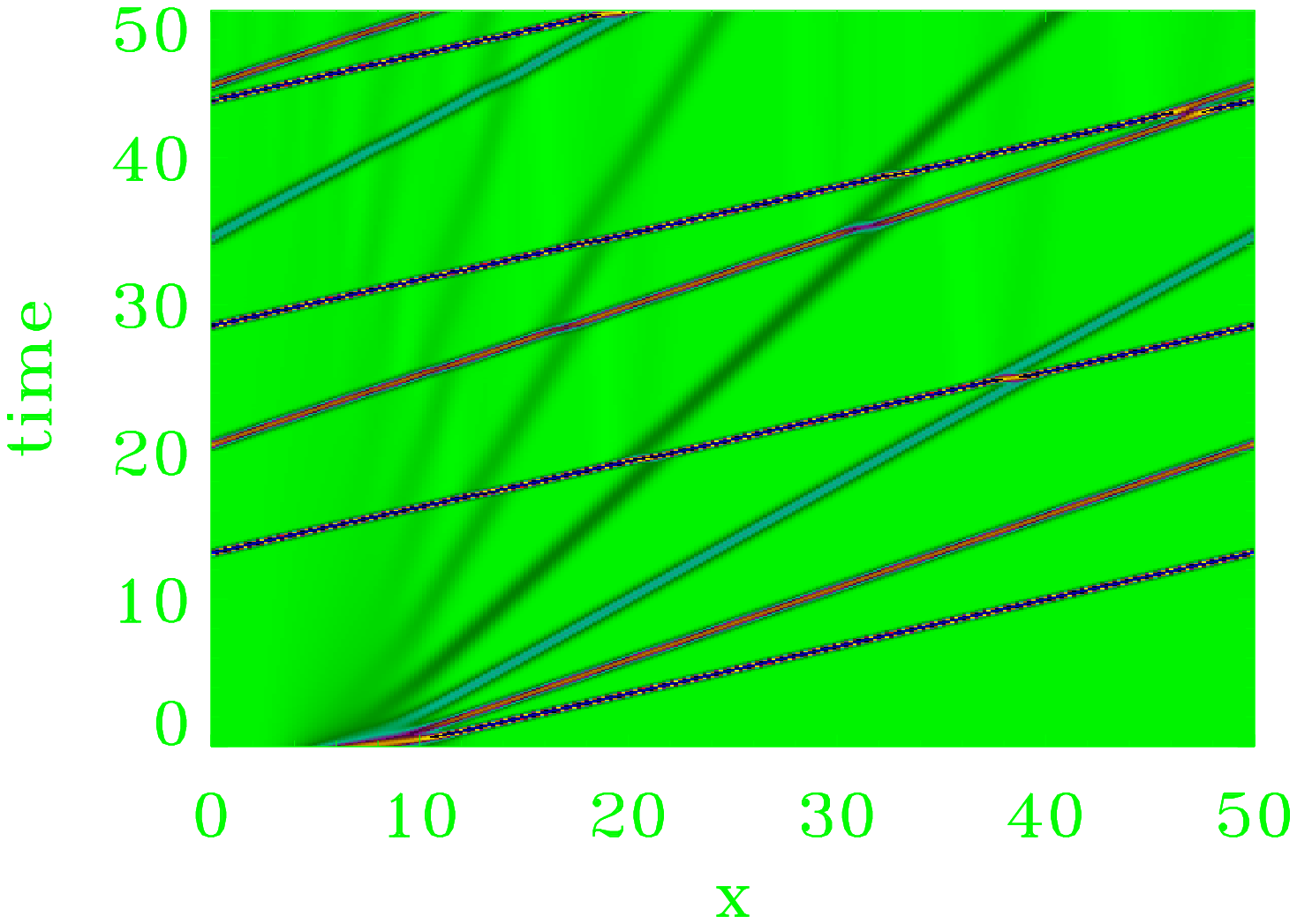}}
\protect\label{01c-w4-cst}
\noindent
Fig. 5
Gray scale contour plot of 
the evolution of the compactons 
in Figs. 4. \\

Then next generic feature is what happens when two compactons of different
speeds collide.  The compactons remain coherent and experience a phase
shift. This is shown in Fig. 6 for the compactons described by Eq. (4.20).\\

Unlike the solitons of the original integrable KdV equation,
these compactons after scattering also can cause ``pair" production
of a compacton and anti compacton. This is shown in Fig. 7 in the resolving
of the wake left behind during the collision shown in Fig. 6.

\epsfxsize=4.5in
\centerline{\epsffile{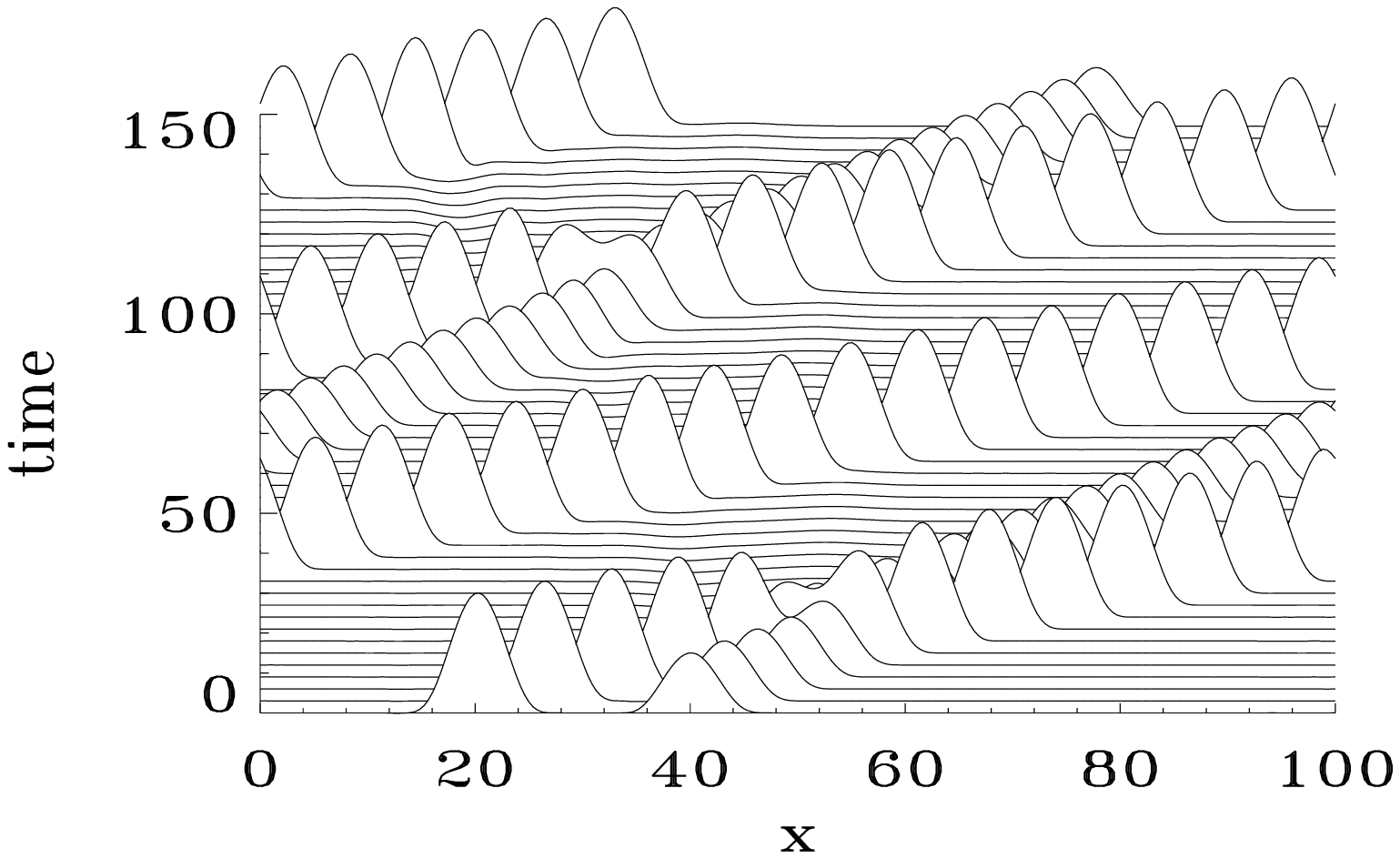}}
\protect\label{01b-col2}
\noindent
Fig. 6. Collision of two solitons for the case
$p=m=n=1, l=0$ and 
$\alpha = 5, \beta = 3, \gamma = 1$.
Two compactons described by Eq. (4.20) collide. One has  speed $c = 2$; 
and the other has speed $c = 1$; 
Note the phase shift 
in the slower compacton 
after the collision.  These 
compactons remained 
coherent, even after dozens 
more collisions.  \\

\epsfxsize=3.5in
\centerline{\epsffile{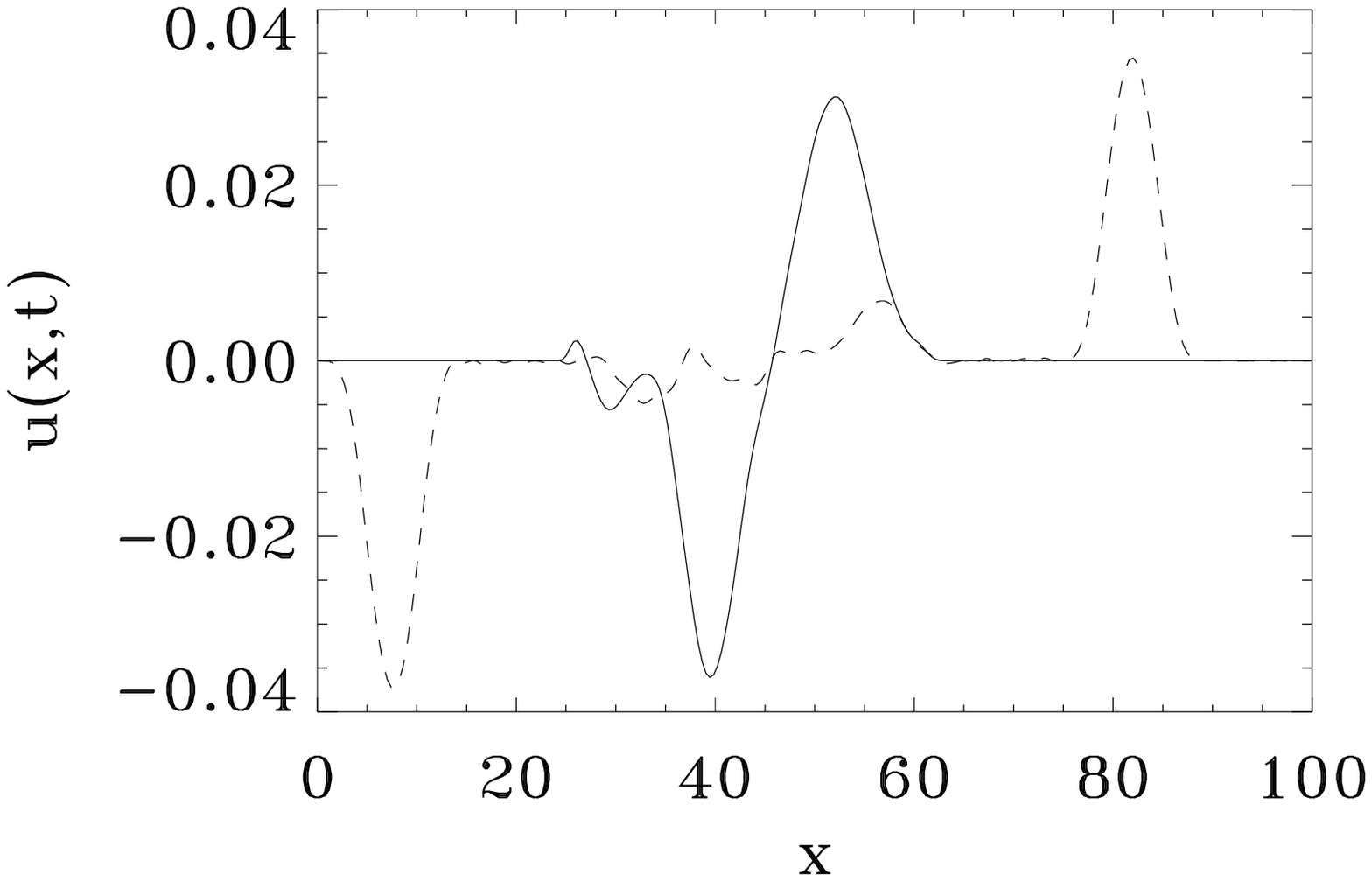}}
\protect\label{01b-col2-rpl}
\noindent
Fig.7.  
A ripple (solid line) is created when 
the compactons first collide in Fig. 6.
This ripple was extracted from the solution at 
$t=25$ and used as an initial condition. 
By time $t=500$, the ripple (dashed line) 
has separated into compactons traveling 
in opposite directions. These compactons 
have a shape proportional to $A \cos^2 ({y \over \sqrt{12}})$. \\

When a compacton and an anti-compacton collide, as well as when one
starts with initial data which is narrower than the width of the stable
compacton one finds numerically blowup at late times. Whether this
is an numerical artifact is not absolutely certain.  This effect
is shown in Fig. 8 for the same compactons as in Figs. 6 and 7. 
\epsfxsize=3.5in
\centerline{\epsffile{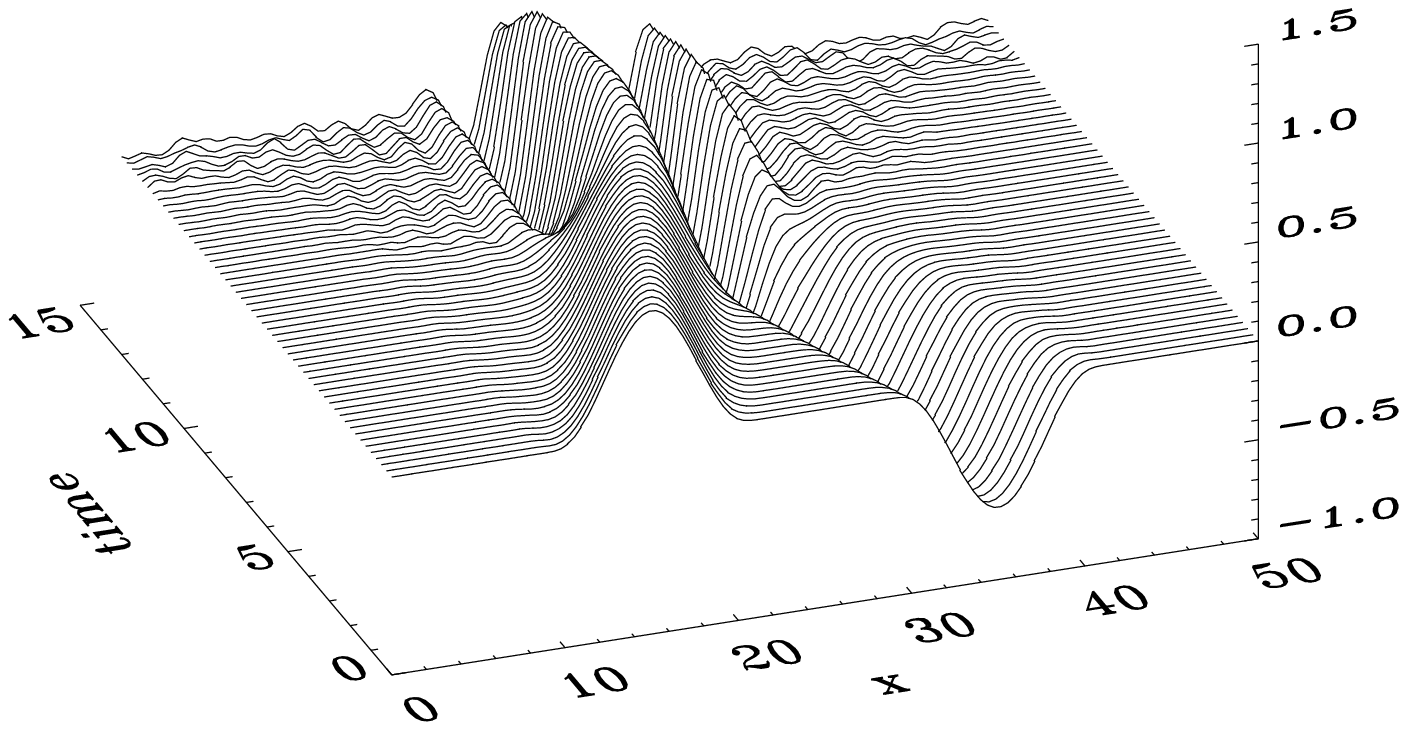}}
\label{01b-updown}
\noindent
Fig. 8. Possible blowup after a compacton anti-compacton collision.
$p=m=n=1, l=0$ and 
$\alpha = 5,\beta = 3, \gamma = 1$.
Two compactons described by Eq. (4.20) with speed $c = +1$ 
and $c = -1$ 
collide.  The numerical solution 
breaks down slightly after 
$t=15$.  It is not clear if the break 
down is due to the steep 
gradients in the solution, or because 
there is a true singularity 
that develops in the equations.

\section*{Appendix- Exact Variational Ansatz}
 In this section we determine to what extent we could recover from the
variational ansatz the exact solitary wave solutions we have discovered earlier 
by trial and error.  That is if we assumed solutions of the form 
\[ A(t) \cos^r
[b(t) (x-q(t))]; {-\pi \over2} \leq b(x-q(t)) \leq {\pi \over 2}\ \]
 for compact
solitary waves and \[ A(t) {\rm sech}^r [b(t) (x-q(t))] \]  for ordinary
solitary waves would we recover all the exact solutions?

First let us show that for the KdV equation and for the generalized KdV equation
that we investigated earlier, we indeed obtain the exact solution. The
Lagrangian for the KdV equation is  
\[
 L= \int~dx~ [ \frac{1}{2} \vp_{x} \vp_{t} - (\vp_{x})^3 - {1 \over 2} 
(\vp_{xx})^{2}].  \] 
The conserved Hamiltonian is given by
\[ H= \int~dx~ 
[(\vp_{x})^3 + {1 \over 2}  (\vp_{xx})^{2}].  \] 
Assuming the trial wave
function: 
\begin{equation} \varphi_x = u(x,t) = A(t) {\rm sech}^2 [b(t) (x-q(t)]
\end{equation} 
we find that the reduced action is 
\begin{equation} 
\Gamma = - P \dot{q} - H[ A(P) ,b],
\end{equation} where \begin{equation} P= \int
{1 \over 2} u^2 dx = {2 A^2(t) \over 3 b} \end{equation}
 and 
\begin{equation}
 H=
{8 \over 15} A^2 ( {2 A \over b} + b ) \end{equation} We can rewrite $H$ in
terms of A as follows: \begin{equation} H= {4 \over 5} P ({ \pm (6bP)^{1/2} +
b^2}) \end{equation} where we have used the two possible solutions: \[ A= \pm
({3b P \over 2})^{1/2} \] Since the Hamiltonian is independent of q,  we have
that  $P$ is conserved.  $b$ is a variable of constraint and is eliminated by
the equation: \begin{equation} {\partial H \over \partial b} =0 = \pm ({6 P
\over b})^{1/2} +2b \end{equation} Only the negative choice for $A$ in terms of
$P$ yields a positive solution for $b$ namely: \begin{equation}
 b= { (6P)^{1/3} \over 4^{2/3}} \end{equation} Eliminating $b$ the reduced
action is \begin{equation} \Gamma= - P \dot{q} + {(3 P) ^{5/3} \over 5}
\end{equation} Varying the action we find the velocity is a constant: \[ \dot{q}
= (3 P) ^{2/3} = c \] Thus $ A= -{c \over 2}$ and we get the usual exact answer:
\begin{equation} u(x,t) = -{c \over 2} {\rm sech}^2[ {c^{1/2} \over 2}(x-ct)]
\end{equation} We also find that  \begin{equation} H= {4 \over 5} P ({ - (6 b
P)^{1/2} + b^2}) \end{equation} has a minimum at the exact value of $b$ for
fixed $P$. 

Next we consider the class of exact compact solitary waves that we found for the
generalized KdV equation of ref. \cite{bib:khare}  .  In this case the
Lagrangian  is  \[
 L= \int~dx~ [ \frac{1}{2} \vp_{x} \vp_{t} + \alpha {1 \over p(p+1)}
(\vp_{x})^{p+2}- \beta  \vp_x ^p (\vp_{xx})^{2}].  \]

and the Hamiltonian is: \[ H =  \int ~dx [-  {\alpha  \over p(p+1)}
(\vp_{x})^{p+2} +\beta  \vp_x ^p (\vp_{xx})^{2}].  \]

Now we assume a solution of the form ($r=2/p$)  \begin{equation} \varphi_x =
u(x,t) = A \cos^{2/p}[d(t)(x-q(t))] \end{equation} We obtain for the reduced 
action \begin{equation} \Gamma = - P \dot{q} - H [A, d] \end{equation} where now
\begin{equation} P ={A^2 \sqrt{\pi} \Gamma(1/2+2/p)  \over 2 d(t) \Gamma(1+2/p)}
\end{equation} \begin{equation}
 H = {{{A^{2 + p}} \{ 4 \beta   {d^2} (2+ 3p +p^2)-\alpha( 4 p 
       + {p^2})  
       \}  {\sqrt{\pi }} 
     { \Gamma}({1\over 2} + {2\over p})}\over 
   {2 d {p^2} \left( 1 + p \right)  
     \left( 2 + p \right)  {\Gamma}(2 + {2\over p})}} \end{equation}

Replacing $A$ by  \[  \left ( {2 d P \Gamma(1+2/p) \over  \sqrt{\pi} \Gamma(1/2
+ 2/p)} \right)^{1/2} \]

we obtain: \begin{eqnarray} H= && (2d)^{p\over 2}\,
   {P}^{1 + {p\over 2}}\, \nonumber \\
 &&  {  \{ 4 \beta   {d^2} (2+ 3p +p^2)-\alpha( 4 p 
       + {p^2})  
       \}  \,
    \Gamma^{{p\over 2}}(1 + {2\over p}) \over 
  \left( 2\,p + 5\,{p^2} + 4\,{p^3} + {p^4} \right) \,
    \pi ^{p\over 4}\,
   \Gamma^{p\over 2}({1\over 2} + {2\over p})} \end{eqnarray}

We determine   the constraint variable $d$ by ${\partial H \over \partial d} =
0$ , we obtain \begin{equation}
 d^2  = {\alpha p^2 \over 4 \beta  (p+1) (p+2)} \end{equation} Lagrange's
equations gives: \begin{equation}
 \dot{q} = c = -{\partial H \over \partial P} = -{p+2 \over 2}  \{{H \over P} \}
\end{equation}

We then have that  \[ A^p  =  {c (p+1) (p+2) \over 2 \alpha } \] and recover our
previous exact result \cite{bib:khare}:

\begin{equation} u(x,t) =  [{ c (p+1) (p+2) \over 2 \alpha}] ^ {1/p} \cos^{2/p}
\{{ p (x - ct) \over  [4 \alpha (p+1) (p+2)]^{1/2}} \} \end{equation}
 As a function of $d$ for fixed $P$, $H$ is a minimum  at the constraint
equation value of $d$. As an example, when $p=1$,$P=1$ and $\beta=1/2$,$\alpha=1$
one obtains for $H[d]$: \[ H=  {2 \over 9} ({d \over 3 \pi})^{1/2} (-5+ 12 d^2)
\] which has a minimum at $d^2 =1/12$.
 
Now let us look at our generalized equation when $\gamma \neq 0$. For the
special case $p=m=n; l=0$ considered in this paper we have that  the Lagrangian
is: \begin{eqnarray}
 L(p=m=n;l=0) &=& \int~dx~ [ \frac{1}{2} \vp_{x} \vp_{t} + \alpha{
{(\vp_{x})^{p+2}} \over {(p+1)(p+2)}} \nonumber \\ &-& \beta (\vp_{x})^{p}
(\vp_{xx})^{2}+ {\gamma \over 2} \vp_x^p  \vp_{xxx}^2  ],\label{eq:kdv2pmn} 
\end{eqnarray} Introducing a trial variational function of the form:
\begin{equation} u = A \cos^{r} d(t) (x-q(t)) \end{equation} with $ r=4/p$   and
the constraint: \begin{equation} \alpha= {(2+r) (4+r) (-4+4 r +r^2) \beta^2
\over \gamma (-2+r^2+ 6 r)^2} \end{equation}

Using the fact that  \[  P = A^2 \sqrt{\pi} {\Gamma(1/2 +r)  \over 2 d  ~ \Gamma
(1+ r)}  \] to eliminate $A$ in favor of $P$, we again find we can write the
reduced action as \begin{equation} \int dt \, \{ - P \dot{q} - H[P,d] \}
\end{equation} where \begin{eqnarray} H&&= (2d) ^{2/r} P^{1 + {2/r}} {r^2}
\Gamma [1 + r]^{1 + {2/r}} \times  \nonumber\\ && ( 12 \beta^2 + 8 \beta  {d^2}
\gamma  - 
       32 {d^4} {{\gamma }^2} + 20 \beta^2  r - 
       32 \beta  {d^2} \gamma  r + 
       176 {d^4} {{\gamma }^2} r \nonumber \\ &&- 19 \beta^2  {r^2} - 
       32 \beta  {d^2} \gamma  {r^2} - 
       172 {d^4} {{\gamma }^2} {r^2} - 24 \beta^2  {r^3} \nonumber \\ && +  152
\beta  {d^2} \gamma  {r^3} - 
       152 {d^4} {{\gamma }^2} {r^3} - 4 \beta^2 {r^4} + 
       50 \beta  {d^2} \gamma  {r^4} \nonumber \\ && - 152 {d^4} {{\gamma }^2}
{r^4} + 
       4 \beta  {d^2} \gamma  {r^5} - 
       40 {d^4} {{\gamma }^2} {r^5} - 
       3 {d^4} {{\gamma }^2} {r^6}  ) \times  \nonumber \\
 &&( {4 \gamma  {{\pi }^{{1/r}}} 
     {{\left( -2 + 6 r + {r^2} \right) }^2} 
   \Gamma[3 + r]   \Gamma[{1/2} + r]}^{2/r} 
      )^{-1}
 \end{eqnarray}

From the equation that eliminates the constraint variable $d$: \[ {\partial H
\over \partial d} =0 \] we find there are two solutions for $d^2$. One solution, 
\begin{equation} d^2 = {\beta \over \gamma (r^2 + 6r -2)} \end{equation} is a
minimum of $H[d]$ for fixed $P$ and is an exact solution of the generalized KdV
equation. The other solution for $d^2$ \begin{equation}
 d^2= {{\beta \,\left( 3 + 2\,r \right) \,
     \left( -4 + 4\,r + {r^2} \right) }\over 
   {\gamma \,\left( -2 + 6\,r + {r^2} \right) \,
     \left( 8 + 4\,r + 3\,{r^2} \right) }} \end{equation} is a maximum of the
energy $H[d]$ for fixed $P$ and is not a solution of the equation of motion.  
An example discussed earlier is the case $p=r=2$ with $\beta=1$, $\gamma=1/14 $,
and $\alpha= 96/7$. In that case we have \begin{equation} H[d] = - { 8 P^2 \over
9 \pi} (10d - 5 d^3 + d^5), \end{equation}
 with two extrema: $d=1$ which is a  minimum of H and  yields an exact
solution, $ u= \sqrt{c/2} \cos^2(x-ct)$,  and $d=\sqrt{2}$ which is a 
maximum and  leads to  $ u= \sqrt{3c/2} \cos^2 \sqrt{2}(x-ct)$ which is not a
solution of the original generalized KdV equation.

\section*{Acknowledgements}

This work was supported in part by the DOE. We would like to  thank Darryl Holm
for useful suggestions. One of us (A.K.) would like to thank Los Alamos National
Laboratory for its hospitality.

\end{document}